\documentclass[conference]{IEEEtran}
\IEEEoverridecommandlockouts
\usepackage{cite}
\usepackage{amsmath,amssymb,amsfonts}
\usepackage{algorithmic}
\usepackage{graphicx}
\usepackage{textcomp}
\usepackage{xcolor}
\usepackage{subcaption}

\def\BibTeX{{\rm B\kern-.05em{\sc i\kern-.025em b}\kern-.08em
    T\kern-.1667em\lower.7ex\hbox{E}\kern-.125emX}}
\begin{document}

\title{A Hybrid Sequential Convex Programming Framework for Unbalanced Three-Phase AC OPF\\
}

\author{\IEEEauthorblockN{Sary Yehia}
\IEEEauthorblockA{\textit{Department of Electrical and Electronic Engineering} \\
\textit{The University of Manchester}\\
Manchester, UK \\
sary.yehia@manchester.ac.uk}
\and
\IEEEauthorblockN{Alessandra Parisio}
\IEEEauthorblockA{\textit{Department of Electrical and Electronic Engineering} \\
\textit{The University of Manchester}\\
Manchester, UK \\
alessandra.parisio@manchester.ac.uk}

}

\maketitle

\begin{abstract}
This paper presents a hybrid Sequential Convex Programming (SCP) framework for solving the unbalanced three-phase AC Optimal Power Flow (OPF) problem. The method combines a fixed McCormick outer approximation of bilinear voltage–current terms, first-order Taylor linearisations, and an adaptive trust-region constraint to preserve feasibility and promote convergence. The resulting formulation remains convex at each iteration and ensures convergence to a stationary point that satisfies the first-order Karush–Kuhn–Tucker (KKT) conditions of the nonlinear OPF. Case studies on standard IEEE feeders and a real low-voltage (LV) network in Cyprus demonstrate high numerical accuracy with optimality gap below $0.1\%$ and up to 2× faster runtimes compared to IPOPT. These results confirm that the method is accurate and computationally efficient for large-scale unbalanced distribution networks.
\end{abstract}


\section{Introduction}
The rapid transformation of power systems poses unprecedented challenges for distribution networks. Unlike transmission grids, distribution systems are inherently unbalanced due to asymmetric line parameters, single-phase loads, and diverse distributed energy resource (DER) connections [1], [2]. These characteristics intensify voltage unbalance and losses, highlighting the need for accurate unbalanced three-phase optimal power flow (OPF) formulations. The AC OPF problem determines optimal operating points subject to physical and operational constraints. However, its nonlinear and nonconvex nature limits global optimality guarantees without specific convex conditions [3]. Unbalanced configurations further amplify this complexity, motivating tractable yet accurate OPF models.

Efforts to address OPF nonconvexity have mainly followed two directions. The first employs convex relaxations, notably semidefinite programming (SDP) [4] and second-order cone programming (SOCP) [5]. While SDP provides tighter bounds, SOCP offers superior scalability. These methods yield outer approximations that can be exact under conditions such as radial topology or balanced operation [6]. However, in unbalanced or meshed systems, mutual couplings often break exactness, producing infeasible or physically inconsistent solutions. The second direction relies on linearisation-based models, which approximate nonlinearities around a reference point. Classical DC OPF neglects reactive power and is unsuitable for distribution networks [7], while first-order Taylor-based formulations [8] enable fast, locally accurate models for real-time control. Yet, their accuracy degrades as operating points deviate from the linearisation point potentially causing divergence [9], [10].

To address the issues of both directions, recent studies have explored sequential linear programming (SLP) and sequential convex programming (SCP) schemes. Early formulations introduced IV-ACOPF and SLP methods [9], later extended through convex–concave procedures to loadability and multi-objective variants [10]. Enhanced SLP algorithms with trust-region and feasibility restoration mechanisms have improved convergence reliability [11], while exact SLP formulations using supporting hyperplanes have achieved KKT convergence in large-scale networks [12]. Nonetheless, these algorithms often incur high iteration costs, limited topology generality, and/or rely on heuristic numerical safeguards rather than physics-informed constraints.

The approaches reviewed above primarily address balanced networks, leaving the challenges of unbalanced three-phase systems largely unexplored. Extensions of SOCP and SDP have incorporated phase coupling [13], inverter control [14], and chordal decomposition for rank reduction [15]. Similarly, linear OPF variants using mixed-integer programming [16], first-order or fixed-point linearisation [17], and ZIP load models [18] have been proposed. Despite these efforts, current methods remain limited by approximation errors, scalability challenges, restrictive topology assumptions, and/or sensitivity to initialisation.

The method proposed in this paper integrates McCormick relaxations with first-order affine linearisations under a trust-region constraint, unifying tractability and feasibility within a sequential convex framework. The method targets unbalanced three-phase AC OPF in both radial and meshed distribution networks, where asymmetry and nonlinearities challenge existing single-strategy approaches. McCormick envelopes convexify bilinear voltage–current terms, while first-order surrogates maintain local accuracy. The adaptive trust-region, based on Euclidean norm error bounds, ensures local convergence and approximation fidelity even under poor initialisation.

The key contributions of this paper are summarised as follows:
\begin{itemize}
\item A hybrid SCP framework that combines McCormick relaxations and first-order affine linearisations, bridging the gap between outer relaxation and local linearisation approaches.
\item An adaptive second-order cone (SOC) trust-region constraint that bounds the deviation between linearised and bilinear terms, yielding a convex bound on the linearisation error and ensuring convergence to KKT-consistent stationary points of the nonlinear AC OPF.
\item Extensive validation on IEEE benchmark feeders and a real low-voltage (LV) distribution network in Cyprus, demonstrating close agreement with nonlinear IPOPT-based OPF solutions while maintaining convexity at every iteration.
\end{itemize}

The remainder of this paper is organised as follows. Section II introduces the unbalanced three-phase network model and AC OPF formulation. Section III presents the proposed hybrid convex method. Section IV reports numerical results, and Section V concludes the paper.

\section{Problem Formulation}{\label{sec:2}}
This section presents the mathematical formulation of the AC OPF problem for unbalanced three-phase distribution networks. The formulation is based on the current injection model (CIM), which provides a compact and general representation of nodal power balances in rectangular coordinates [19]. This representation captures the effects of unbalanced operation. The resulting optimisation problem is nonlinear and nonconvex, motivating the convexification and approximation methods developed later. 

\subsection{Network Model}

Consider a network represented by the set of buses $\mathcal{N}$ and the set of branches $\mathcal{E}$. A subset of buses hosts generators, denoted by $\mathcal{G}$. Each bus $i \in \mathcal{N}$ may accommodate up to three phases $\phi \in \{a,b,c\}$. Let $N_{\text{ph}}$ denote the total number of bus-phase nodes, i.e., the total number of pairs $(i,\phi)$ such that phase $\phi$ exists at bus $i$. At each bus-phase node, the complex voltage and current are represented in rectangular coordinates as $v_{i,\phi} = V^{R}_{i,\phi} + j V^{I}_{i,\phi}$ and $i_{i,\phi} = I^{R}_{i,\phi} + j I^{I}_{i,\phi}$, where $(\cdot)^{R}$ and $(\cdot)^{I}$ indicate the real and imaginary parts, respectively.

The complex power contributions of generators and loads at each bus-phase node are defined as 
$S_{i,\phi}^{G} = P_{i,\phi}^{G} + j Q_{i,\phi}^{G}$ and 
$S_{i,\phi}^{D} = P_{i,\phi}^{D} + j Q_{i,\phi}^{D}$, 
with $P$ and $Q$ representing active and reactive power terms.

To capture the network interactions, all bus-phase voltages and currents are stacked into vectors 
$\mathbf{V} \in \mathbb{C}^{N_{\text{ph}}}$ and $\mathbf{I} \in \mathbb{C}^{N_{\text{ph}}}$. 
These quantities are coupled through the nodal form of Ohm's law:
\begin{equation}
    \mathbf{I} = Y \mathbf{V},
\end{equation}
where $Y \in \mathbb{C}^{N_{\text{ph}} \times N_{\text{ph}}}$ is the bus admittance matrix. Unless otherwise specified, all voltage--current products 
(e.g., \(VI\), \(V^R I^R\), etc.) are defined component-wise 
across all bus--phase nodes, while vector and matrix operations 
(e.g., \(\mathbf{I} = Y \mathbf{V}\)) follow standard linear algebra conventions. 

Each block of $Y$ is obtained from the per-branch $3 \times 3$ admittance matrices, which capture self and mutual impedances, 
transformers, tap-changing regulators, and shunt elements. Then, the $(i,j)$ block of $Y$ is given by
\begin{equation}
Y_{ij} =
\begin{cases}
    \displaystyle \sum_{m : (i,m)\in \mathcal{E}} y_{im}, & i = j, \\[1.2em]
    -y_{ij}, & i \neq j, \\[0.5em]
    0, & \text{otherwise},
\end{cases}
\end{equation}
where $y_{ij} \in \mathbb{C}^{3\times 3}$ denotes the branch admittance matrix between buses $i$ and $j$, 
with $(i,j) \in \mathcal{E}$.

The complex power injection at each bus-phase node $(i,\phi)$ is given by
\begin{equation}
    S_{i,\phi} = P_{i,\phi} + \mathrm{j} Q_{i,\phi} 
    = v_{i,\phi}\,(i_{i,\phi})^{*} \label{eq:power_def},
\end{equation}
where $(\cdot)^{*}$ denotes the complex conjugate.

\subsection{AC OPF Formulation}
The objective of the AC OPF problem can represent different operational goals such as cost minimisation, loss reduction, or voltage profile improvement. 
In this work, we focus on improving the voltage profile by penalizing deviations of bus voltages from their nominal complex values. 
The resulting equation for this penalty is defined as
\begin{equation}
    f_{\mathbf{V}} 
    = 
    (V_{i,\phi}^R - \widehat{V}_{\phi}^R)^2 
    + 
    (V_{i,\phi}^I - \widehat{V}_{\phi}^I)^2,
    \quad 
    \begin{cases}
        i \in \mathcal{N},\\
        \phi \in \{a,b,c\}
    \end{cases}
    \label{eq:fv_corrected}
\end{equation}
where 
$\widehat{V}_{\phi}^R$ and $\widehat{V}_{\phi}^I$ are the reference real and imaginary components of the nominal phasor voltages.
 
Then, the complete AC OPF can be formulated as a nonlinear programming problem:
\begin{align}
    \min_{\mathbf{V}, \mathbf{I}, P^G, Q^G} \quad 
        & \sum_{i \in \mathcal{N}} \sum_{\phi \in \{a,b,c\}} f_{\mathbf{V}} \label{eq:acopf_obj}\\[0.5em]
    \text{s.t.} \quad 
        & (1), (3) \nonumber \\[0.3em]
        & P_{i,\phi} = P_{i,\phi}^G - P_{i,\phi}^D,
          \quad Q_{i,\phi} = Q_{i,\phi}^G - Q_{i,\phi}^D 
          \label{eq:acopf_pq}\\[0.3em]
        & \underline{P}_{i,\phi}^{G} \leq P_{i,\phi}^G \leq \overline{P}_{i,\phi}^{G},
          \quad \underline{Q}_{i,\phi}^{G} \leq Q_{i,\phi}^G \leq \overline{Q}_{i,\phi}^{G}
          \label{eq:acopf_PQ_limits}\\[0.3em]
        & \underline{V}_{i,\phi} \leq |V_{i,\phi}| \leq \overline{V}_{i,\phi} \label{eq:v_limits}
\end{align}
for all $i \in \mathcal{N}$ and $\phi \in \{a,b,c\}$.
Constraint \eqref{eq:acopf_pq} represents the active and reactive 
power balance at each bus-phase node. Constraint \eqref{eq:acopf_PQ_limits} imposes the operating limits of the generators by bounding 
their active and reactive outputs within feasible capability ranges. Finally, 
constraint \eqref{eq:v_limits} enforces the engineering limits of the voltage magnitudes. \\
\indent The nonlinear coupling between voltages and currents through the power definition \eqref{eq:power_def} is the fundamental source of nonconvexity in the AC OPF problem. This is evident when we expand \eqref{eq:power_def} in rectangular coordinates
\begin{align}
    P_{i,\phi} &= V_{i,\phi}^{R} I_{i,\phi}^{R} + V_{i,\phi}^{I} I_{i,\phi}^{I} \label{eq:nonconvex_P}, \\
    Q_{i,\phi} &= V_{i,\phi}^{I} I_{i,\phi}^{R} - V_{i,\phi}^{R} I_{i,\phi}^{I}. \label{eq:nonconvex_Q}
\end{align}
This nonconvexity will be addressed in the approach presented next in Section~\ref{sec:III}.

\section{Hybrid SCP Model} \label{sec:III}
\begin{figure}[t]
    \centering
    \includegraphics[width=0.9\linewidth]{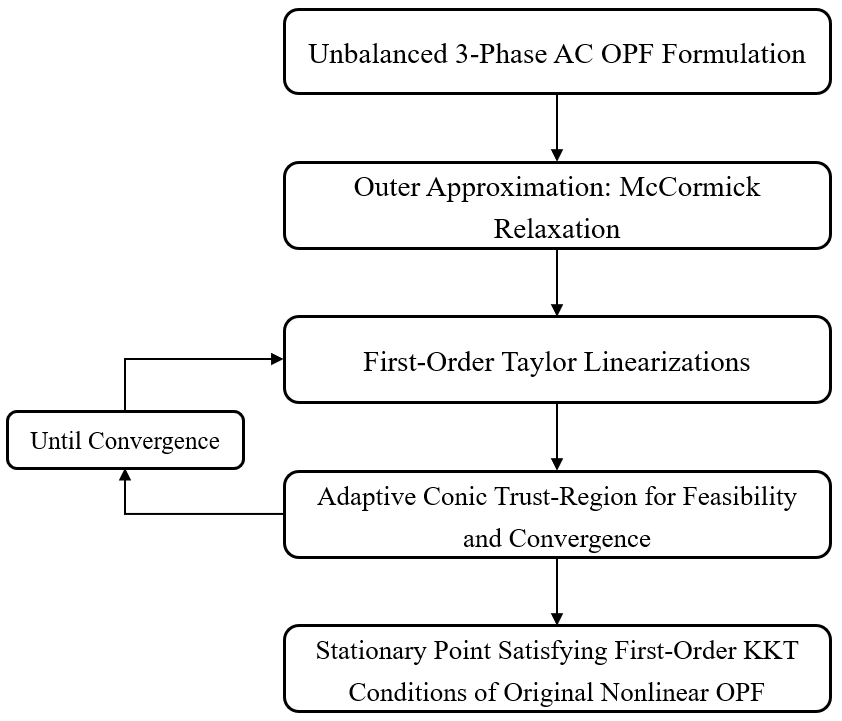}
    \caption{Proposed Hybrid SCP framework for unbalanced three-phase AC OPF.}
    \label{fig:flowchart}
\end{figure}

The nonlinearity in the aforementioned OPF problem comes from \eqref{eq:nonconvex_P} and \eqref{eq:nonconvex_Q}. It introduces bilinear voltage-current terms which make the feasible set of the OPF problem nonconvex. Hence, this section presents an iterative convex algorithm for AC OPF problems. The method first constructs a convex outer approximation of the bilinear voltage-current products using McCormick envelopes. Then, the first-order Taylor approximations are used to locally linearise around an operating point. Lastly, the adaptive conic trust-region constraint ensures validity and convergence. The approach is presented in Fig.~\ref{fig:flowchart} and successively refines the approximation of the original nonconvex OPF until convergence. 

\subsection{Outer Approximation of Bilinear Terms} \label{sec:III-A}
To convexify bilinear products between voltage and current components, the proposed method employs McCormick relaxations. Auxiliary variables are introduced for these products, replacing the exact equalities with convex linear envelopes that form an outer approximation of the bilinear manifold. 

Consider a generic bilinear term $z = xy$, 
where $x$ and $y$ are continuous variables bounded by 
$x \in [\underline{x}, \overline{x}]$ and $y \in [\underline{y}, \overline{y}]$. The McCormick envelope defines the convex hull of all feasible $(x,y,z)$ triplets 
over these bounds using the following four linear inequalities:
\begin{equation}
\begin{split}
    z &\geq \underline{x}y + \underline{y}x - \underline{x}\,\underline{y}, 
    \quad z \geq \overline{x}y + \overline{y}x - \overline{x}\,\overline{y},\\
    z &\leq \overline{x}y + \underline{y}x - \overline{x}\,\underline{y},
    \quad z \leq \underline{x}y + \overline{y}x - \underline{x}\,\overline{y}.
\end{split}
\label{eq:mccormick}
\end{equation}

These four constraints define a polyhedral region that encloses the bilinear manifold 
$z = xy$, as shown in Fig.~\ref{fig:mccormick}. They form a convex outer approximation that includes all feasible points on the true manifold as well as relaxed points in its interior. The tightness of the relaxation depends on the variable bounds where narrower intervals yield stronger approximations.

In the proposed framework, the McCormick region is constructed once using fixed global bounds to avoid the overhead of iterative bound tightening. Accuracy and feasibility recovery are achieved instead via the adaptive trust-region and sequential affine surrogate updates described in Section~\ref{sec: III-C}.

In the OPF formulation, McCormick relaxations are applied systematically to the four
bilinear products of voltage and current components present in \eqref{eq:nonconvex_P} and \eqref{eq:nonconvex_Q}:
\begin{equation}
\begin{split}
    m^{RR} &\approx V^R I^R, \quad 
    m^{RI} \approx V^R I^I,\\
    m^{IR} &\approx V^I I^R, \quad 
    m^{II} \approx V^I I^I,
\end{split}
\label{eq:mc_terms}
\end{equation}
where ``$\approx$'' indicates that the equality is relaxed through the McCormick
envelope \eqref{eq:mccormick}. Note that subscripts $i$ and $\phi$ for buses and phases, respectively, were omitted for clarity. These auxiliary variables then define the active and reactive power injections at each bus--phase node as:
\begin{align}
    P &= m^{RR} + m^{II} \\
    Q &= m^{IR} - m^{RI}.
\end{align}

This convexification replaces each bilinear relation with a set of linear constraints that bound its feasible region. 

\subsection{First-Order Linearisation of Bilinear Terms} \label{sec:III-B}
Although the McCormick relaxation introduced in Section~\ref{sec:III-A} provides a convex outer approximation of the bilinear terms, the resulting feasible region may remain loose and contain nonphysical points due to the relaxation gap. To refine the approximation and guide the solution toward the nonlinear manifold, each bilinear term is locally linearised around the current operating point $(V^{(k)},I^{(k)})$ using a first-order Taylor expansion. This expansion is applied individually to each bilinear voltage--current product:
\begin{align}
    X^{RR} &= V^{R,(k)} I^R + V^R I^{R,(k)} - V^{R,(k)} I^{R,(k)} \label{eq:lin1} \\
    X^{RI} &= V^{R,(k)} I^I + V^R I^{I,(k)} - V^{R,(k)} I^{I,(k)} \label{eq:lin2} \\
    X^{IR} &= V^{I,(k)} I^R + V^I I^{R,(k)} - V^{I,(k)} I^{R,(k)} \label{eq:lin3} \\
    X^{II} &= V^{I,(k)} I^I + V^I I^{I,(k)} - V^{I,(k)} I^{I,(k)} \label{eq:lin4}.
\end{align}

These affine functions exactly match the bilinear term at $(V^{(k)},I^{(k)})$ and provides 
a locally accurate linear surrogate within a small neighbourhood. This linearisation 
captures the local behaviour of the bilinear term and serves as a building block 
for the proposed iterative model.

These relations serve as affine definitions of the local first-order 
linearisations of the original bilinear terms $V^R I^R$, $V^R I^I$, $V^I I^R$, and $V^I I^I$. By replacing the nonlinear equalities with these linearised counterparts, the algorithm obtains a convex subproblem that is solved efficiently at each iteration. \\
\indent The linearisation guarantees that the linearised model is exact at the current iterate 
and locally accurate in its neighbourhood. However, while these first-order surrogates
capture local behaviour, they are not inherently coupled to the McCormick relaxation, 
which spans a larger convex outer set. As a result, the two representations may deviate. 
To address this, Section~\ref{sec: III-C} introduces a physics-informed constraint that limits the deviation between each auxiliary variable and its linearised value. Rather 
than constraining the step size directly, this mechanism anchors the relaxation to its 
local tangent approximation. Thus, it guides the iterates progressively toward the true 
nonlinear manifold as the algorithm advances.

\begin{figure}[t]
    \centering
    \includegraphics[width=0.9\linewidth]{Mc.jpg}
    \caption{McCormick relaxation of the bilinear term $z = x \cdot y$ showing 
    the true surface, convex envelope, and optimal solution.}
    \label{fig:mccormick}
\end{figure}

\subsection{Trust-Region Constraint} \label{sec: III-C}

\subsubsection{Alignment-Based SOC Constraint}
The first-order linearisations introduced in Section~\ref{sec:III-B} provide locally 
accurate affine surrogates for the bilinear voltage--current products. However, 
since the McCormick relaxation defines a broader convex feasible region, 
the auxiliary variables $m^{RR}, m^{RI}, m^{IR}, m^{II}$ may deviate from 
their linearised counterparts, particularly across successive iterations. 
To ensure consistency between these two model layers, we introduce a 
trust-region style coupling constraint that limits the deviation between each 
McCormick auxiliary variable and its corresponding linear surrogate. \\
\indent This deviation is quantified through a quadratic trust-region measure:
\begin{align}
    &(m^{RR} - X^{RR})^2 + (m^{RI} - X^{RI})^2  \notag \\
    &\quad + (m^{IR} - X^{IR})^2 + (m^{II} - X^{II})^2 \;\leq\; \delta^2,
    \label{eq:prox_quad}
\end{align}
where $\delta$ denotes the radius of the trust-region ball centred on the local surrogates $X$. 

For solver compatibility and numerical robustness, this quadratic inequality 
is reformulated as an SOC constraint, which preserves 
convexity and can be handled efficiently by modern conic optimisation solvers. 
Defining the deviation vector $e$ at iteration $k$ as
\begin{equation}
    e = \begin{bmatrix}
        m^{RR} - X^{RR} \\
        m^{RI} - X^{RI} \\
        m^{IR} - X^{IR} \\
        m^{II} - X^{II}
    \end{bmatrix},
\end{equation}
the SOC representation is expressed as
\begin{equation}
    \|e\|_2 \leq \delta.
\end{equation}

Consequently, each iteration solves for a point that is simultaneously feasible 
with respect to the convex relaxation and consistent with the local linearisation. 
Unlike conventional trust-region methods that constrain the step size of primal variables (e.g., voltages or currents), the proposed trust-region operates on the auxiliary variables of the convex relaxation. This alignment-based method directly bounds the deviation between the relaxation and linearisation subspaces. This ensures that the affine linearisations \eqref{eq:lin1}--\eqref{eq:lin4} remain physically meaningful and that their approximation quality is preserved throughout the iterative process.

\subsubsection{Adaptive Adjustment Strategy} \label{sec:III-C-2}
To promote convergence, the trust-region radius, denoted $\delta$, is adaptively updated across iterations. This adaptive mechanism allows the algorithm to tolerate large mismatch during early stages, while progressively tightening the trust-region radius to improve consistency between the convex relaxation and local linear models as the solution stabilises.

\begin{figure*}[t]
    \centering
    \begin{subfigure}[t]{0.25\textwidth}
        \centering
        \includegraphics[width=\textwidth]{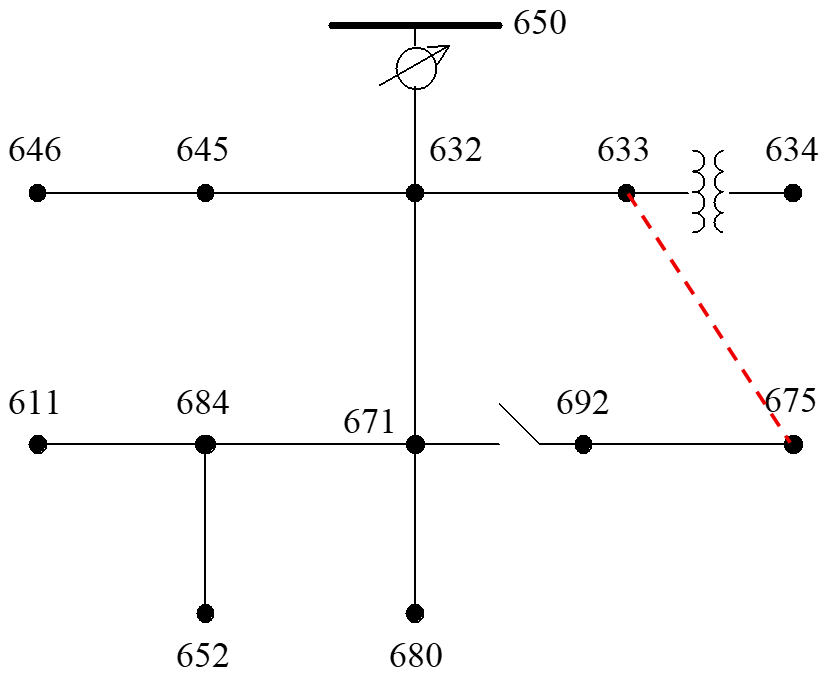}
        \caption{13-bus system (meshed).}
        \label{fig:ieee13}
    \end{subfigure}
    \hfill
    \begin{subfigure}[t]{0.38\textwidth}
        \centering
        \includegraphics[width=\textwidth]{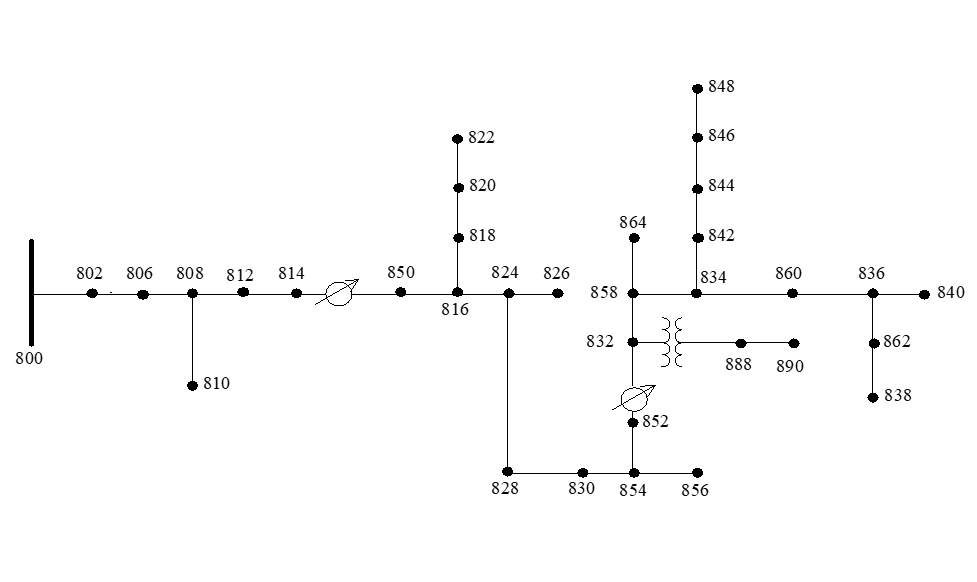}
        \caption{34-bus system.}
        \label{fig:ieee34}
    \end{subfigure}
    \hfill
    \begin{subfigure}[t]{0.34\textwidth}
        \centering
        \includegraphics[width=\textwidth]{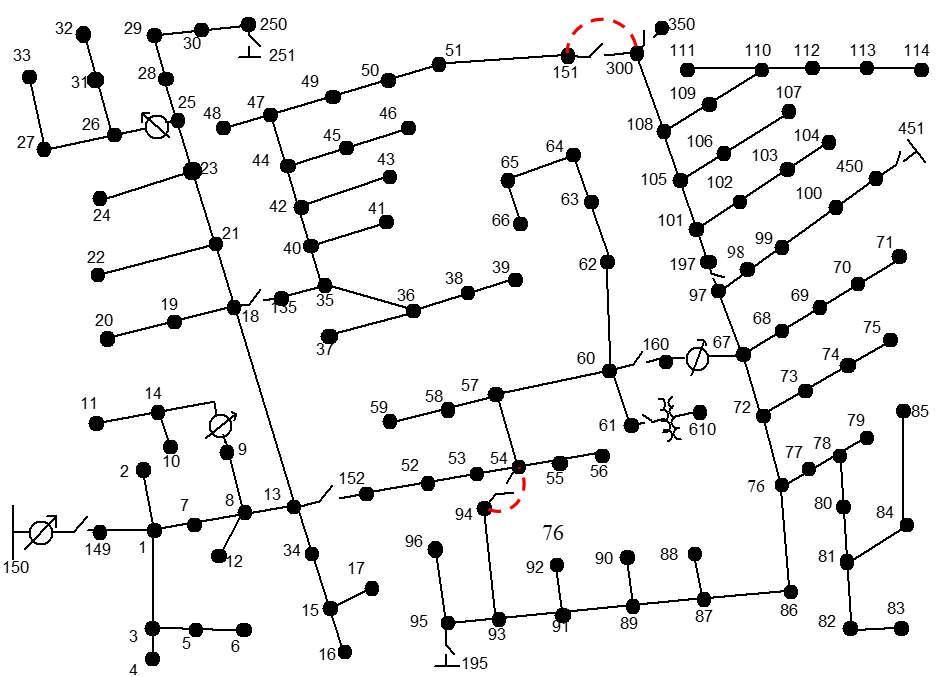}
        \caption{123-bus system (meshed).}
        \label{fig:ieee123}
    \end{subfigure}
    \caption{IEEE distribution test systems used for evaluation.}
    \label{fig:test-systems}
\end{figure*}

The adjustment process is guided by a voltage deviation metric, defined as
\begin{equation}
\Delta v^{(k)} = 
\max
\begin{aligned}[t]
& \big| V^{R,(k)} - V^{R,(k-1)} \big| \\
& \quad + \big| V^{I,(k)} - V^{I,(k-1)} \big|
\end{aligned}
\label{eq:delta_v}
\end{equation}
which quantifies the maximum step-to-step change in the real and imaginary voltage components.

When the deviation $\Delta v^{(k)}$ falls below a threshold $\tau$, the ball radius is contracted according to
\begin{equation}
\delta^2 \leftarrow \max \big(\underline{\delta}^2, \, \alpha \delta^2 \big),
\quad 0 < \alpha < 1,
\label{eq:shrink}
\end{equation}
where $\underline{\delta}^2$ denotes the minimum allowable radius. Conversely, if $\Delta v^{(k)}$ exceeds the threshold, the radius is expanded as
\begin{equation}
\delta^2 \leftarrow \min \big(\overline{\delta}^2, \, \beta \delta^2 \big),
\quad \beta \geq 1,
\label{eq:expand}
\end{equation}
with $\overline{\delta}^2$ defining the maximum allowable radius.

This mechanism achieves a balance between convergence and accuracy. A larger region in the early iterations provides flexibility to recover feasibility even under poor initialisations, while progressive contraction ensures alignment with the nonlinear manifold and promotes convergence toward a physically meaningful solution.

\subsection{Convergence Analysis}
Following the manifold-based linearisation theory of [20] which defines a bounded neighbourhood for valid affine power-flow approximations, the proposed analysis interprets the adaptive trust-region as a practical mechanism enforcing this local validity. Convergence toward a feasible solution of the nonlinear AC OPF is examined through the relationship between the auxiliary variables ${m}$, their linearised surrogates ${X}$, and the true nonlinear bilinear terms ${V}{I}$.

In Euclidean norm, the deviation between an auxiliary variable ${m}$ and the exact bilinear product ${V}{I}$ can be bounded using the triangle inequality as:
\begin{equation}
\|m - VI\| \le 
\underbrace{\|m - X\|}_{\text{(a)}} +
\underbrace{\|X - VI\|}_{\text{(b)}},
\label{eq:triangle}
\end{equation}
isolating two error sources.

\paragraph{Relaxation Deviation}
From the alignment-based SOC constraint, each iteration satisfies
\begin{equation}
\|m - X\| \;\leq\; {\delta},
\label{eq:prox_enforce}
\end{equation}
bounding the distance between the McCormick variable $m$ 
and its affine surrogate $X$. This ensures that 
$m$ remains confined within a convex ball of radius $\delta$, 
maintaining coherence between the relaxed and linearised representations.

\paragraph{Linearisation Error}
The local affine surrogate is defined as
\begin{equation}
X \;=\; V^{(k)} I \;+\; V I^{(k)} \;-\; V^{(k)} I^{(k)}.
\end{equation}
Subtracting the exact bilinear term $VI$ and taking the norm on each side yields
\begin{equation}
\|X - VI\| \;=\; \|(V - V^{(k)})(I - I^{(k)})\|.
\end{equation}

This deviation depends on the proximity of the current iterate \((V^{(k)}, I^{(k)})\) to the true nonlinear quantities \((V, I)\). 
As the algorithm converges, i.e., \(V^{(k)} \to V\) and \(I^{(k)} \to I\), the linearisation error term vanishes, ensuring that the surrogate \(X\) coincides with the true bilinear product \(VI\). 

Combining both terms (a) and (b) gives
\begin{equation}
\|m - VI\| \;\leq\; \delta \;+\; \|(V - V^{(k)})(I - I^{(k)})\|. 
\label{eq:conc}
\end{equation}
As the trust-region radius shrinks $\delta \to 0$, and the iterates stabilise $V^{(k)} \to V, \; I^{(k)} \to I$, both error terms vanish ensuring that $m \to VI$. The relaxed problem thus aligns with the nonlinear manifold, and the solution converges to a feasible, near-exact OPF point.

Convergence is declared when both the trust-region radius $\delta$ which bounds the bilinear model mismatch,
and the voltage deviation metric $\Delta v$ which reflects the step size in the state space fall below their respective thresholds. This guarantees alignment between the convex relaxation and the true nonlinear solution.

\section{Numerical Evaluation}
This section assesses the performance of the proposed hybrid SCP framework in terms of convergence behaviour, solution accuracy, and computational efficiency. The methodology is validated on standard IEEE radial distribution networks, their meshed variants obtained via selective loop-forming adjustments, and a realistic symmetric LV distribution feeder from Cyprus. Together, these benchmarks capture both canonical test cases and practical operating conditions. This enables a comprehensive evaluation of the algorithm under diverse topological structures and loading scenarios.

\subsection{Simulation Setup}
All simulations are implemented in MATLAB R2024b using the YALMIP modeling framework. Convex subproblems are solved with GUROBI, while nonlinear AC OPF benchmarks are obtained via IPOPT to evaluate solution accuracy.  

Experiments are executed on a laptop equipped with an Intel\textsuperscript{\textregistered} Core\texttrademark{} i5-1335U (2.40 GHz) processor and 16 GB RAM. All variables are expressed in per-unit (p.u.), with line parameters, voltage limits, and complex power injections derived from IEEE test feeders and field measurements from the Cyprus LV network.

A flat start for voltages is used. The trust-region radius is initialised at a moderately high value $\delta^2 = 10^{-1}$ to promote early flexibility and is adaptively tightened according to the strategy described in Section~\ref{sec:III-C-2}. Convergence is declared when both $\Delta v < 10^{-3}$ and $\delta^2 < 10^{-6}$, ensuring stable voltage iterates and consistent alignment between linearised and relaxed representations.

\begin{table*}[!t]
\renewcommand{\arraystretch}{1.3}
\centering
\caption{Comprehensive Performance and Accuracy Comparison Between IPOPT and Hybrid SCP}
\label{table_combined}
\resizebox{\textwidth}{!}{%
\begin{tabular}{|c|c|c|c|c|c|c|c|}
\hline
\textbf{Test Case} & 
\textbf{IPOPT Time (s)} & 
\textbf{Hybrid SCP Time (s)} & 
\textbf{Iterations} & 
\textbf{Speedup Ratio} & 
\textbf{Optimality Gap (\%)} & 
\textbf{Max Voltage Error (p.u.)} & 
\textbf{Mean Voltage Error (p.u.)} \\
\hline
IEEE 13-bus & 1.499 & 0.869 & 3 & 1.725 & $0.00167$ & $5.52\times10^{-5}$ & $1.09\times10^{-5}$ \\
IEEE 13-bus (meshed) & 1.897 & 1.380 & 3 & 1.375 & $0.00697$ & $1.283\times10^{-5}$ & $6.508\times10^{-6}$ \\
IEEE 34-bus & 2.079 & 1.067 & 4 & 1.948 & $0.00554$ & $5.628\times10^{-7}$ & $5.148\times10^{-7}$ \\
IEEE 123-bus & 2.093 & 1.368 & 5 & 2.162 & $0.0599$ & $1.253\times10^{-4}$ & $2.141\times10^{-4}$ \\
IEEE 123-bus (meshed) & 3.276 & 2.680 & 6 & 1.222 & $0.0255$ & $1.174\times10^{-4}$ & $7.611\times10^{-5}$ \\
Cyprus 241-bus & 3.896 & 1.736 & 4 & 2.244 & $0.0333$ & $2.641\times10^{-7}$ & $1.92\times10^{-7}$ \\
\hline
\end{tabular}%
}
\end{table*}

\begin{figure}[b]
    \centering
    \includegraphics[width=0.48\textwidth]{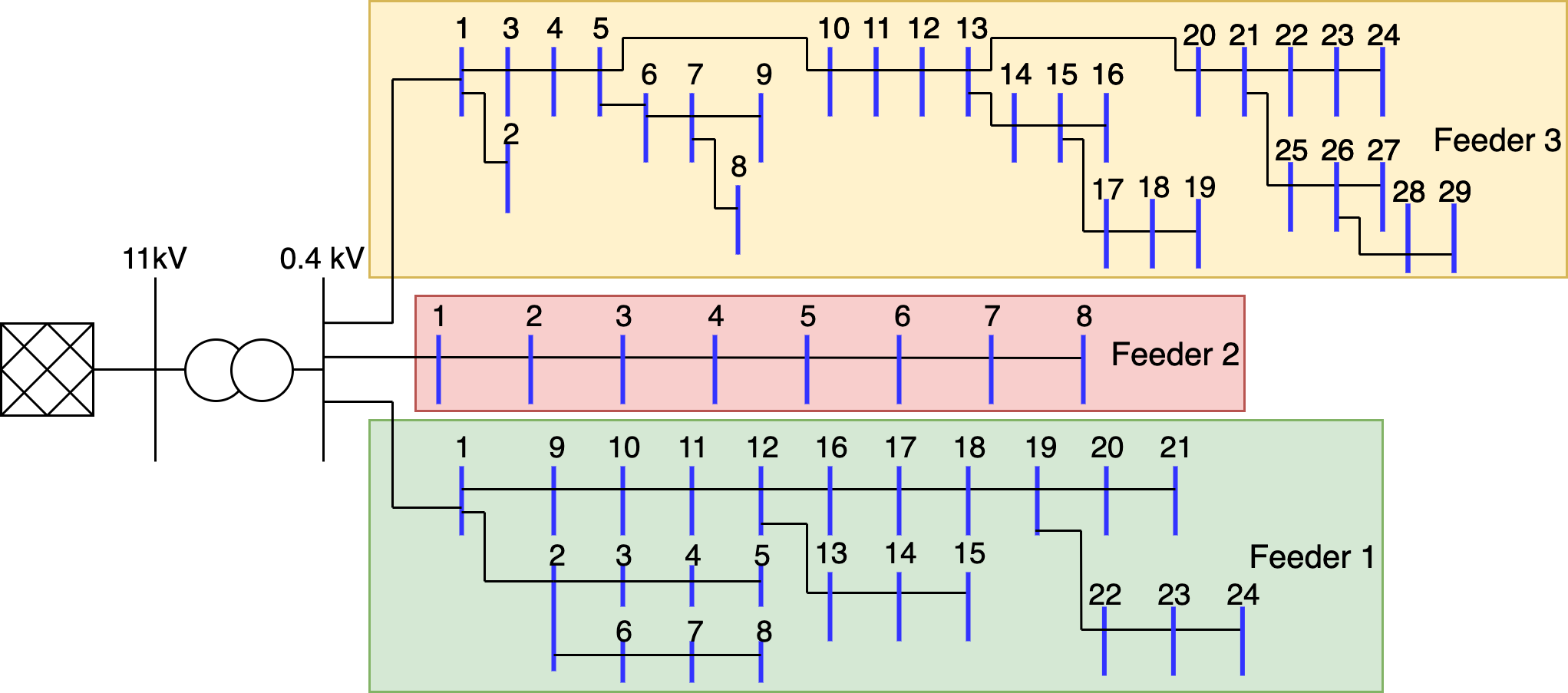}
    \caption{Single-line diagram of the Cyprus LV distribution feeder.}
    \label{fig:cyprus_lv_network}
\end{figure}

\subsection{Test Systems}
\subsubsection{Standard IEEE Test Feeders}

The proposed framework is evaluated on three standard IEEE distribution feeders -- the 13-bus, 34-bus, and 123-bus systems [21] -- to assess scalability and performance across different network sizes.
All feeders are modelled as unbalanced three-phase networks with asymmetric line impedances, phase-specific loads, and voltage-dependent elements where applicable. These systems vary in topology, load diversity, and phase configuration, providing a representative range from small to large-scale radial systems.
Each network includes distributed loads and enforces voltage constraints at all buses. Substation is modelled as a slack bus with fixed voltage magnitude 1.0 p.u. and zero phase angle. No distributed generation is included in the baseline to isolate solver performance on load-only systems.

\subsubsection{IEEE Test Feeders under Meshed Configurations}
To evaluate performance under non-radial topologies, two IEEE feeders, 13- and 123-bus systems, are modified to form a meshed configuration by introducing a few tie-lines between non-adjacent buses [22]. These additions create independent loops while preserving overall sparsity.
The meshed variants test the method’s generality beyond strictly radial structures, where additional voltage–current coupling increases problem dimensionality and sensitivity to initialisation.
Fig.~\ref{fig:test-systems} illustrates both configurations, with added tie-lines shown in dashed red.

\subsubsection{Cyprus LV Distribution System}
To assess real-world applicability, a 241-bus Cypriot LV feeder is considered, as shown in Fig.~\ref{fig:cyprus_lv_network} [23]. The network, modelled in \texttt{pandapower}, includes a 315~kVA, 11~kV/0.4~kV Dyn transformer feeding three radial feeders with distinct structures and load conditions:
\begin{itemize}
\item Feeder 1: 24 nodes, 72 loads
\item Feeder 2: 8 nodes, 48 loads
\item Feeder 3: 29 nodes, 58 loads
\end{itemize}

The original test case specifies identical 0.4~kW / 0.132~kVAr residential loads with power factor = 0.95. To introduce realistic heterogeneity and unbalance, loads are randomised across buses: active powers $P \in [0.2,0.8]$~kW, while power factors are drawn from $0.92$–$0.98$, resulting in reactive powers within $Q \in [0.041,0.34]$~kVAr. 

All feeders use standard LV cable parameters and are modelled as unbalanced three-phase systems. The resulting dataset provides a high-fidelity benchmark combining phase asymmetry, spatial load diversity, and network unbalance, enabling rigorous validation of the proposed SCP-based OPF framework.

\subsection{Results and Discussion}
Table~\ref{table_combined} summarises the comparative performance of the proposed hybrid SCP approach against the nonlinear solver IPOPT. The proposed method consistently outperforms IPOPT in computational efficiency, achieving speedup ratios between 1.22 and 2.24 across all test systems. For smaller feeders such as the IEEE~13- and 34-bus networks, SCP converges within 3--4 iterations and reduces computation time by nearly 50\%. For larger networks IEEE~123- and Cyprus~241-bus, the method maintains similar accuracy with only 4--6 iterations, reducing total runtime by more than 30\%. These results demonstrate that the hybrid SCP framework scales effectively with system size while preserving numerical stability.

The optimality gap between SCP and IPOPT solutions remains below $0.1\%$ in all cases, confirming that the convexified subproblems do not compromise global solution quality. The maximum and mean voltage errors, typically between $10^{-5}$ and $10^{-7}$~p.u., highlight the excellent voltage fidelity achieved by the SCP algorithm. These negligible discrepancies indicate that the local linearisations retain sufficient accuracy to approximate the nonlinear OPF solution effectively.

Across all test systems, the hybrid SCP algorithm exhibits monotonic convergence with decreasing trust-region radius and voltage deviations $\Delta v$. Convergence is typically achieved within 3--6 iterations, consistent with the adaptive radius contraction strategy described in Section~III-C. The evolution of $\Delta v$ confirms that feasibility is preserved throughout iterations, demonstrating that the method can converge reliably even from infeasible initial points.

In the meshed network configurations, the SCP method maintains versatile performance, with slightly lower speedup ratios 1.22--1.38 due to denser nodal coupling and reduced sparsity in the linearised subproblems. Nevertheless, convergence remains within 3--6 iterations, and the optimality gap stays on the same order of magnitude as in the radial cases. This confirms that the SCP framework effectively handles both radial and weakly meshed networks without significant loss of accuracy or stability.

Overall, the results demonstrate that the proposed hybrid SCP approach achieves a balance between computational efficiency and accuracy. It attains near-identical solutions to the nonlinear IPOPT benchmark while providing substantial runtime reductions, making it well-suited for real-time and large-scale OPF applications.

\section{Conclusion}
This paper presented a hybrid SCP method for unbalanced three-phase AC OPF in distribution networks. By integrating McCormick relaxations, first-order affine linearisations, and an adaptive conic trust-region, the proposed approach forms a convex approximation that maintains feasibility while accurately capturing nonlinear AC power flow relations. Case studies on the IEEE 13-, 34-, and 123-bus feeders and a real LV Cypriot network showed near-exact accuracy to the nonlinear IPOPT benchmark, with optimality gap below 0.1\%. The method converged in a few iterations, achieved over 2× runtime speedup, and converged under diverse topologies.

Despite these advantages, the approach remains a local method and may require mild parameter tuning in large or ill-conditioned systems to ensure fast convergence. At this stage, the work primarily focuses on the formulation and validation of the OPF problem. Future work will extend this framework into a full OPF simulation environment handling dynamic scenarios. In addition, the method will be generalised to multi-energy networks and integrated with predictive control architectures to enable coordinated and real-time operation.

\section*{Acknowledgment}
This project has received funding from European Union's Framework Programme for Research and Innovation Europe Horizon Europe (HORIZON) Marie Skłodowska-Curie Actions Doctoral Networks (MSCA-DN) under the Grant Agreement No. 101120278. The authors would also like to thank Dr. Petros Aristidou and Mr. Savvas Panagi from the Cyprus University of Technology for providing real operational data from the Cyprus distribution network.


\begin{thebibliography}{1}
\bibitem{b1} ‘Review of distribution network phase unbalance: Scale, causes, consequences, solutions, and future research direction’, CSEE J. Power Energy Syst., 2020, doi: 10.17775/CSEEJPES.2019.03280.
\bibitem{b2} I. A. Ibrahim and M. J. Hossain, ‘Low Voltage Distribution Networks Modeling and Unbalanced (Optimal) Power Flow: A Comprehensive Review’, IEEE Access, vol. 9, pp. 143026–143084, 2021, doi: 10.1109/ACCESS.2021.3120803.
\bibitem{b3} S. Gopinath et al., ‘Proving global optimality of ACOPF solutions’, Electr. Power Syst. Res., vol. 189, p. 106688, Dec. 2020, doi: 10.1016/j.epsr.2020.106688.
\bibitem{b4} X. Bai, H. Wei, K. Fujisawa, and Y. Wang, ‘Semidefinite programming for optimal power flow problems’, Int. J. Electr. Power Energy Syst., vol. 30, no. 6–7, pp. 383–392, Jul. 2008, doi: 10.1016/j.ijepes.2007.12.003.
\bibitem{b5} R. A. Jabr, ‘Radial Distribution Load Flow Using Conic Programming’, IEEE Trans. Power Syst., vol. 21, no. 3, pp. 1458–1459, Aug. 2006, doi: 10.1109/TPWRS.2006.879234.
\bibitem{b6} S. H. Low, ‘Convex Relaxation of Optimal Power Flow—Part II: Exactness’, IEEE Trans. Control Netw. Syst., vol. 1, no. 2, pp. 177–189, Jun. 2014, doi: 10.1109/TCNS.2014.2323634.
\bibitem{b7} A. G. Bakirtzis and P. N. Biskas, ‘A decentralized solution to the DC-OPF of interconnected power systems’, IEEE Trans. Power Syst., vol. 18, no. 3, pp. 1007–1013, Aug. 2003, doi: 10.1109/TPWRS.2003.814853.
\bibitem{b8} Z. Yang, H. Zhong, Q. Xia, and C. Kang, ‘Solving OPF using linear approximations: fundamental analysis and numerical demonstration’, IET Gener. Transm. Distrib., vol. 11, no. 17, pp. 4115–4125, Nov. 2017, doi: 10.1049/iet-gtd.2017.1078.
\bibitem{b9} A. Castillo, P. Lipka, J.-P. Watson, S. S. Oren, and R. P. O’Neill, ‘A successive linear programming approach to solving the IV-ACOPF’, IEEE Trans. Power Syst., vol. 31, no. 4, pp. 2752–2763, Jul. 2016, doi: 10.1109/TPWRS.2015.2487042.
\bibitem{b10} L. P. M. I. Sampath, B. V. Patil, H. B. Gooi, J. M. Maciejowski, and K. V. Ling, ‘A trust-region based sequential linear programming approach for AC optimal power flow problems’, Electr. Power Syst. Res., vol. 165, pp. 134–143, Dec. 2018, doi: 10.1016/j.epsr.2018.09.002.
\bibitem{b11} W. Wei, J. Wang, N. Li, and S. Mei, ‘Optimal Power Flow of Radial Networks and Its Variations: A Sequential Convex Optimization Approach’, IEEE Trans. Smart Grid, vol. 8, no. 6, pp. 2974–2987, Nov. 2017, doi: 10.1109/TSG.2017.2684183.
\bibitem{b12} S. Mhanna and P. Mancarella, ‘An Exact Sequential Linear Programming Algorithm for the Optimal Power Flow Problem’, IEEE Trans. Power Syst., vol. 37, no. 1, pp. 666–679, Jan. 2022, doi: 10.1109/TPWRS.2021.3097066.
\bibitem{b13} M. M.-U.-T. Chowdhury, B. D. Biswas, and S. Kamalasadan, ‘Second-Order Cone Programming (SOCP) Model for Three Phase Optimal Power Flow (OPF) in Active Distribution Networks’, IEEE Trans. Smart Grid, vol. 14, no. 5, pp. 3732–3743, Sep. 2023, doi: 10.1109/TSG.2023.3241216.
\bibitem{b14} M. He et al., ‘A SOCP-Based ACOPF for Operational Scheduling of Three-Phase Unbalanced Distribution Systems and Coordination of PV Smart Inverters’, IEEE Trans. Power Syst., vol. 39, no. 1, pp. 229–244, Jan. 2024, doi: 10.1109/TPWRS.2023.3236881.
\bibitem{b15} W. Wang and N. Yu, ‘Chordal Conversion Based Convex Iteration Algorithm for Three-Phase Optimal Power Flow Problems’, IEEE Trans. Power Syst., vol. 33, no. 2, pp. 1603–1613, Mar. 2018, doi: 10.1109/TPWRS.2017.2735942.
\bibitem{b16} J. S. Giraldo, P. P. Vergara, J. C. Lopez, P. H. Nguyen, and N. G. Paterakis, ‘A Linear AC-OPF Formulation for Unbalanced Distribution Networks’, IEEE Trans. Ind. Appl., vol. 57, no. 5, pp. 4462–4472, Sep. 2021, doi: 10.1109/TIA.2021.3085799.
\bibitem{b17} K. Girigoudar and L. A. Roald, ‘Linearized Three-Phase Optimal Power Flow Models for Distribution Grids with Voltage Unbalance’, in 2021 60th IEEE Conference on Decision and Control (CDC), Austin, TX, USA: IEEE, Dec. 2021, pp. 4214–4221. doi: 10.1109/CDC45484.2021.9683780.
\bibitem{b18} Tsinghua University et al., ‘Linear three-phase power flow for unbalanced active distribution networks with PV nodes’, CSEE J. Power Energy Syst., vol. 3, no. 3, pp. 321–324, Oct. 2017, doi: 10.17775/CSEEJPES.2017.00240.
\bibitem{b19} G. Ferro, M. Robba, D. D’Achiardi, R. Haider, and A. M. Annaswamy, ‘A distributed approach to the Optimal Power Flow problem for unbalanced and mesh networks’, IFAC-Pap., vol. 53, no. 2, pp. 13287–13292, 2020, doi: 10.1016/j.ifacol.2020.12.159.
\bibitem{b20} S. Bolognani and F. Dorfler, ‘Fast power system analysis via implicit linearization of the power flow manifold’, in 2015 53rd Annual Allerton Conference on Communication, Control, and Computing (Allerton), Monticello, IL: IEEE, Sep. 2015, pp. 402–409. doi: 10.1109/ALLERTON.2015.7447032.
\bibitem{b21} W. H. Kersting, ‘Radial distribution test feeders’, IEEE Trans. Power Syst., vol. 6, no. 3, pp. 975–985, Aug. 1991, doi: 10.1109/59.119237.
\bibitem{b22} J. Huang, B. Cui, X. Zhou, and A. Bernstein, ‘A Generalized LinDistFlow Model for Power Flow Analysis’, in 2021 60th IEEE Conference on Decision and Control (CDC), Austin, TX, USA: IEEE, Dec. 2021, pp. 3493–3500. doi: 10.1109/CDC45484.2021.9682997.
\bibitem{b23} S. Panagi, ‘SPS-L/Cyprus-Grids: Initial Release’. Zenodo, Sep. 30, 2025. doi: 10.5281/zenodo.17232114.


\end{thebibliography}
\end{document}